\documentclass[12pt]{article}
\usepackage{latexsym,amsmath,amssymb,epsfig,graphicx}

\newcommand{\pkt}{\; .}
\newcommand{\kma}{\; ,}

\newcommand{\bea}{\begin{eqnarray}}
\newcommand{\eea}{\end{eqnarray}}
\newcommand{\be}{\begin{equation}}
\newcommand{\ee}{\end{equation}}
\newcommand{\beast}{\begin{eqnarray*}}
\newcommand{\eeast}{\end{eqnarray*}}
\newcommand{\eqn}[1]{(\ref{#1})}

\newcommand{\call}{{\cal L}}
\newcommand{\calf}{{\cal F}}

\newcommand{\calm}{{\cal M}}

\newcommand{\calv}{{\cal V}}

\newcommand{\bfk}{{\bf k}}
\newcommand{\bfx}{{\bf x}}
\newcommand{\bfy}{{\bf y}}

\renewcommand{\Re}{{\rm Re}\,}
\renewcommand{\Im}{{\rm Im\,}}

\begin{document}

\begin{titlepage}
\noindent
DESY09-161   \\
 hep-th/yymmnn\\
October 2009

\vspace{8mm}
\begin{center}
{\Large \bf
Initial time singularities and admissible initial states 
for a system of coupled scalar fields}
\\\vspace{8mm}
{\large  J\"urgen Baacke\footnote{e-mail:~
juergen.baacke@tu-dortmund.de}} \\
{  Fakult\"at Physik, Technische Universit\"at Dortmund \\
D - 44221 Dortmund, Germany
}\\
\vspace{4mm}
and\\
\vspace{4mm}
{\large  Nina Kevlishvili\footnote{e-mail:~
nina.kevlishvili@desy.de}} \\
{ Deutsches Elektronen Synchrotron DESY \\ D - 22603 Hamburg,
Germany, \\
Andronikashvili Institute of Physics, GAS\\
 0177 Tbilisi, Georgia
}\\
\vspace{5mm}

\bf{Abstract}
\end{center}

We discuss the problem of initial states for a system of coupled scalar
fields out of equilibrium in
the one-loop approximation. The fields consist of classical
background fields, taken constant in space, and quantum fluctuations.
If the initial state is the adiabatic vacuum, i.e.,  the ground state
of a Fock space of particle excitations that diagonalize the mass matrix,
the energy-momentum tensor is infinite at $t=0$, its most singular
part  behaves as $1/t$. When the system is coupled to gravity this presents
a problem that we solve by a Bogoliubov transformation of the naive
initial state. As a side result we also discuss the canonical formalism 
and the adiabatic particle number for such a system. Most of the
formalism is presented for Minkowksi space. Embedding the
system and its dynamics into a flat FRW universe is straightforward
and we briefly address the  essential modifications.
\end{titlepage}


\setcounter{page}{2}
\section{Introduction}\label{introduction}
The question of initial states in nonequilibrium quantum field
theory has found considerable interest recently
\cite{Maslov:1997bf,Baacke:1997zz,Anderson:2001th,
Goldstein:2002fc,Danielsson:2002mb,
Burgess:2003hw,
Schalm:2004qk,Anderson:2005hi,Collins:2005nu,Collins:2006bg,
Greene:2004np,Borsanyi:2008ar}, on various grounds.
As  a very practical aspect it was realized in numerical 
simulations of quantum fields in cosmology 
\cite{Baacke:1997rs,Baacke:1997zz,Baacke:1999nq}, that
the energy-momentum tensor had initial time  singularities if the
initial state was taken as the naive Fock-space vacuum. These
had to be removed when coupling the field to gravity, e.g. in a Friedmann
universe. A more speculative aspect that has attracted some
interest recently was the question, whether the choice of initial state
can be expected to leave an imprint in the CMB spectrum. In any case
it is a question of principle, to what extent the choice of initial state
is constrained in an interacting theory of particles by consistency
requirements.

As a general aspect of
quantum field theory, the problem of initial conditions was realized
long ago. It ultimately can traced back to the fact that 
one switches on the interaction at some time $t=0$. In the case of 
nonequilibrium dynamics we have to impose initial conditions
for the background fields. In most applications the
initial state was taken to be the adiabatic vacuum, which would
be the ground state if all background fields were held fixed forever.
When one starts to evolve the system dynamically at least 
the second derivative
of the fields will be discontinuous on account of the
second order differential equation, which for $t>0$ 
determines their evolution. In cosmology it is already
the first derivative of the scale parameter which for
$t>0$ is determined by the Friedmann equation. 
Such singularities  have been noted
for the first time by St\"uckelberg \cite{Stueckelberg:1951zz},
they are discussed briefly in the textbook of Bogoliubov and Shirkov
\cite{Bogoliubov:1980}. The phenomenon has  been identified
as a kind of  `Casimir effect' connected to
the initial time surface by Symanzik \cite{Symanzik:1981wd}.
In the context of quantum field theory out of equilibrium
the presence of such singularities has been noticed by
various authors.

The solution proposed by Symanzik is the introduction of surface
counterterms in addition to the usual ``bulk'' counterterms
of perturbative quantum field theory. In the context of nonequilibrium
quantum field theory this line has been pursued in Refs.
\cite{Collins:2005nu,Collins:2006bg}.

The introduction of initial time surfaces singles out the particular
time at which one starts the evolution to the extent that 
the surface counterterms become part of the field theory for $t > 0$. 
The approach of modifying the naive initial vacuum state
\cite{Cooper:1987pt,Baacke:1997zz,Anderson:2005hi}
seems to be more pragmatic; the idea is to find the minimal
requirements on an initial state that could  arise
from a previous dynamical evolution. The latter aspect is discussed
in \cite{Baacke:1999ia}. The technique used for constructing
such an initial state consists of finding a Bogoliubov
transformation of the naive adiabatic vacuum.  
For the single-channel case and scalar fluctuations this has been done in 
Ref. \cite{Baacke:1997zz} and for fermion fluctuations
 in Ref. \cite{Baacke:1998di}.
These results were used in Refs. \cite{Baacke:1999gc,Baacke:1999nq} 
in formulating the renormalized equations in a flat FRW universe.

Our approach is based on a mode function formalism that has been
introduced, for coupled channels, in Ref. \cite{Baacke:2003bt}.
The formalism ensures the conservation of energy
with one-loop or Hartree quantum backreaction and has been renormalized
along the lines of Ref. \cite{Baacke:1996se}.
There the initial state was chosen to be the adiabatic vacuum
based on a Fock space of particle excitations that diagonalize the
initial mass matrix. It is this initial state that we will improve
here. In Ref. \cite{Baacke:2003bt} renormalization is based on a 
perturbative expansion close
to standard perturbation theory, in the same way as in Ref.
\cite{Baacke:1996se}. The same expansion was used
in Ref. \cite{Baacke:1997zz} for analyzing the initial time singularity. This
analysis can be carried over in a straightforward way 
to the case of coupled fields. 
Most other analyses of the singular early time behavior
were based on the eikonal expansion. An eikonal formalism for
coupled systems has been formulated recently \cite{Nilles:2001fg}.
We are not aware, however, of an eikonal expansion
for coupled-channel systems. 
 
The quantum expansion of the fluctuation fields
is formulated in such a way that the canonical commutation 
relations hold for $t=0$. It was not discussed in Ref. \cite{Baacke:2003bt}
how they continue to hold for $t>0$. Though this is to be expected
it is not entirely obvious, and in fact leads some nontrivial 
relations for the fluctuation modes which prove to be useful for our
formalism. This is discussed in Appendix \ref{canform}. 
Another approach to the canonical formalism for coupled-channel
systems was introduced in Ref. \cite{Nilles:2001fg}, and this is
another reason for verifying that the scheme
of Ref. \cite{Baacke:2003bt} works correctly. 

Though our main subject here is the choice of the initial state,
with hindsight of coupling the system to gravity, we take the occasion
for discussing the concept of adiabatic particle number within
our formalism. This is suggested by the fact that we have to discuss
Bogoliubov transformations for coupled systems anyway and that particle
numbers are usually defined by the coefficients of these transformations.

The paper is organized as follows:
In Sec. \ref{model} we introduce the model we want to consider,
a system of two coupled quantum fields with masses and a general fourth
order potential. We define the decomposition into classical and
fluctuation fields and the evolution of the fluctuations.
In Sec. \ref{initialstate} we 
discuss the behavior of the Green's function at early times
and construct a Bogoliubov 
transformation in order to reduce
the leading singular behavior such that the leading time derivatives 
become finite at the intial time.
In Sec. \ref{energymomentumtensor} we present the 
expectation value of the energy-momentum tensor in the Bogoliubov-transformed
initial state. In Sec. \ref{particles} we discuss the concept of adiabatic
particle number.
Some more technical subjects are transferred to the Appendices: the Bogoliubov 
transformation for coupled systems in Appendix \ref{bogotrans}
and some aspects of the canonical formalism in Appendix \ref{canform}.


\section{The model} \label{model}
\setcounter{equation}{0}
We consider a system of two coupled scalar quantum fields
with a Lagrangian density of the form
\be
\call=\left[\frac12 \partial_\mu \phi_i\partial
^\mu \phi_i +\frac 12 m_i^2 \phi_i^2\right] + 
\frac{\lambda_{ij}}{4}\phi_i^2\phi_j^2
\kma\ee
where the indices $i,j$ take the values $1$ and $2$.
The hybrid model of inflation \cite{Linde:1993cn,
Copeland:1994vg,Garcia-Bellido:1997wm}
with
\be
\call=\frac12 \partial_\mu \phi\partial
^\mu \phi+\frac12 \partial_\mu \chi\partial^\mu \chi  +\frac 12 m^2 \phi^2 +
\frac\alpha4\left(\chi^2-v^2\right)^2
+\frac\lambda2\phi^2\chi^2
\ee
is of this form with $\lambda_{11}=0$,$\lambda_{12}=\lambda$,
$\lambda_{22}=\alpha$,$m_1^2= m^2$ and $m_2^2=-\alpha v^2$.
Also some models involving supersymmetric flat directions \cite{Olive:2006uw,
Allahverdi:2006xh} are of this type. The generalization to a 
general mass matrix and a general fourth
order potential is possible, but we do not want do overburden
the formalism with a profusion of indices. Also, the limitation to two fields
is not essential. 

We separate the fields $\phi$ into classical fields and
fluctuations
via
\be 
\phi_i=\varphi_i+\psi_i
\pkt\ee
The classical Lagrangian density then retains the form
\be
\call^{(0)}=\left[\frac12 \partial_\mu \varphi_i\partial
^\mu \varphi_i +\frac 12 m_i^2 \varphi_i^2\right] + 
\frac{\lambda_{ij}}{4}\varphi_i^2\varphi_j^2
\kma \ee
while the fluctuation Lagrangian, of second order in the fluctuations, becomes
\be
\call^{(2)}=\left[\frac12 \partial_\mu \psi_i\partial
^\mu \psi_i +\frac 12 m_i^2 \psi_i^2\right] 
+ 
\frac{\lambda_{ij}}{2}\left[\varphi_i^2\psi_j^2
+2 \varphi_i\varphi_j\psi_i\psi_j\right]
\pkt\ee
This can  be written as
\be
\call^{(2)}=\sum_{i=1}^2\left[\frac12 \partial_\mu \psi_i\partial
^\mu \psi_i +\frac 12 \calm^2_{ij}(\varphi) \psi_i\psi_j\right] 
\kma\ee
with
\bea\nonumber
\calm^2_{11}&=&m^2_1 + 3\lambda_{11}\varphi_1^2
+\lambda_{12}\varphi_2^2 \kma
\\
\calm_{12}^2&=& 2\lambda_{12}\varphi_1\varphi_2
\kma\\
\calm^2_{22}&=&m^2_2 + 3\lambda_{22}\varphi_2^2
+\lambda_{12}\varphi_1^2
\pkt\eea
If the field is coupled to gravity in a flat FRW universe,
the fluctuation mass matrix takes a similar form.
After conformal rescaling of fields and momenta (see, e.g., Refs.
\cite{Birrell:1982,Baacke:1997rs}) one just
has to replace
\be
m_i^2 \to \left[m_i^2+(\xi_i-\frac16)R\right]a^2 
\pkt\ee
Here $a$ is the scale parameter, $R$ the Ricci scalar, and
the $ \xi_i$ are the conformal couplings. 

In the following
we restrict ourselves to homogeneous background fields $\varphi_i(t)$, so the
mass matrix depends on time only.
In the FRW universe the time parameter is conformal time, and we have an
additional time dependence via $a(\tau)$ and $R(\tau)$.

We separate the fluctuation mass matrix into its initial
value and a ''potential'' $\calv$ via
\be\label{calvdef}
\calm^2_{ij}(t)=\calm^2_{ij}(0)+\calv_{ij}(t)
\pkt
\ee
We diagonalize the initial mass matrix by
\be\label{diag0}
\calm^2_{ij}(0)f_{j0}^\alpha=m^2_{\alpha0}f^\alpha_{i0}
\pkt\ee
The eigenvectors $f^\alpha_{i0}$ are chosen to be real, and normalized
to unity:
\be\label{norm0}
\sum_{i=1}^2f^\alpha_{i0}f^\beta_{i0}=\delta^{\alpha\beta}
\pkt \ee
The latin subscripts refer to the field components, as before, and the 
Greek superscripts 
refer to the two independent solutions of the eigenvalue equation.
We now define a set of mode functions $f_i^\alpha(k,t)$ for 
homogeneous background
field in the following way:
\\
(i) their time evolution is determined by
\be\label{fdgl}
\ddot f^\alpha_i(k,t)+k^2 f^\alpha_i(k,t)+
\calm^2_{ij}(t)f^\alpha_j(k,t)=0
\,\,;\ee
(ii) the initial conditions are specified as
\bea \label{finit}
f^\alpha_i(k,0)&=&f^\alpha_{i0} \kma
\\\label{fpinit}
\dot f^\alpha_i(k,0)&=&-i \Omega_{\alpha0} f^\alpha_{i0}
\kma\eea
where we have introduced the frequencies
\be
\Omega_{\alpha0}(k)=\sqrt{m^2_{\alpha0}+k^2}
\pkt
\ee
The functions $f_i^\alpha(k,t)$ form a set of 
linearly independent solutions of the
system of mode equations.

The fields $\psi_i(\bfx,t)$ are quantum fields. For a homogeneous
background we can expand them as
\be\label{fieldexp0}
\psi_i(\bfx,t)=\sum_\alpha
\int \frac{d^3 k}{(2\pi)^32\Omega_{\alpha0}}
\left[a_\alpha(\bfk)f^\alpha_i(k,t)
+a^\dagger_\alpha(-\bfk)f^{\alpha *}_i(k,t)\right]e^{i\bfk\bfx}
\pkt\ee
The canonical commutation relations are
\be
\left[a_\alpha(\bfk),a^\dagger_\beta(\bfk')\right]
= (2\pi)^3 2 \Omega_{\alpha0}(k)\delta_{\alpha\beta}\delta^3(\bfk-\bfk')
\pkt\ee
In the following we will need the two-point functions 
at the coincidence limit, the ``fluctuation integrals'' 
\bea \nonumber
\calf_{ij}(t)&=&< 0| \psi_i(\bfx,t)\psi_j(\bfx,t)|0 >
\\\label{flucintdef}
&=&\sum_\alpha
\int\frac{d^3k}{(2\pi)^3 2 \Omega_{\alpha0}(k)}
f_i^\alpha(k,t)f_j^{\alpha*}(k,t)
\pkt
\eea
Here the expectation value is taken in the vacuum state of a
Fock space, whose quanta have the initial masses  $m_\alpha(0)$.
This is the ``adiabatic vacuum'', defined by
\be
a_\alpha(\bfk)|0> =0\hspace{10mm} \forall \,\alpha,\bfk
\pkt\ee
Of course this is not the ground state of the system, and the
creation and annihilation operators $a^\dagger_\alpha(\bfk)$
and $a_\alpha(\bfk)$ do not create free particles with the masses
$m_i$. 
We discuss some aspects of the canonical formalism in Appendix \ref{canform};
in particular we establish that the fluctuation integral as defined 
above is real and symmetric in $i$ and $j$,
though this is not apparent on the right hand side
of Eq \eqn{flucintdef}.


\section{The initial time singularity of the Green's function and the
modified initial state}
\setcounter{equation}{0} \label{initialstate}

The quantum backreaction of the fluctuations onto the classical
fields can be derived using  the
closed-time-path formalism \cite{Schwinger:1960qe,Keldysh:1964ud}. 
For the quantum field
theories that we consider here, it has been formulated in several
seminal publications 
\cite{Calzetta:1986ey,Jordan:1986ug,Ringwald:1987ui,Cooper:1987pt,Boyanovsky:1992vi}. 
We do  not repeat this here. 
If one just considers the  one-loop quantum backreaction 
the relevant equations take a rather intuitive form. The 
equations of motion for the classical fields become
\bea\nonumber
&&\ddot\varphi_1+m_1^2 \varphi_1 + \lambda_{11}\varphi_1^3
+\lambda_{12}\varphi_1\varphi_2^2
\\
&&\hspace{20mm}+3\lambda_{11}\varphi_1\calf_{11}
+\lambda_{12}\varphi_1\calf_{22}
+2\lambda_{12}\varphi_2\calf_{12}=0
\kma\eea 
and an analogous equation for $\varphi_2$.
As will be analyzed below, the fluctuation integrals $\calf_{ij}(t)$
are singular at $t=0$, the time where we start the evolution.
As $t \searrow 0$ it behaves as $t\ln t$. Though this represents
a mathematical singularity, it is finite and even zero
at $t=0$. So it  will not
prevent us from starting a numerical simulation.
The singular behavior becomes a problem when we couple the
field to gravity. The dynamics of the FRW scale factor $a$
is determined by the energy-momentum tensor, which
involves second time derivatives of the two-point function. 
If one analyzes the energy-momentum
tensor, one indeed finds, near $t=0$, a time dependence of 
the form $1/t$ in $T^\mu_\mu$. This then prevents one from starting the 
dynamical evolution. Of course, even in flat space this infinity
is an undesirable and unphysical feature of the energy-momentum
tensor.

As the energy-momentum tensor is a rather involved expression, especially
after renormalization, we first consider the fluctuation integral
and find a way to remove its initial singularity, such that
its first and second time derivatives at $t=0$ become finite.
This requires less algebra and, as we have seen previously 
\cite{Baacke:1997zz},  this is sufficient for making the energy-momentum
tensor finite near $t=0$.

The fluctuation integrals are ultraviolet divergent. The
divergences can be analyzed \cite{Baacke:1996se} by expanding with respect to
orders in $\calv$ which is equivalent to expanding with respect
to the couplings $\lambda_{ij}$. This allows one to remove the divergent
parts and the dynamics is determined by the
remaining finite parts. A closer analysis shows that, on the level
of fluctuation integrals, the contributions of zeroth and first order 
in $\calv$
are ultraviolet divergent. One finds
\cite{Baacke:1996se}, up to first order in $\calv$, 
\bea\nonumber \label{fluctint}
&&\calf_{ij}=
\int\frac{d^3 k}{(2\pi)^3}\sum_\alpha\frac{1}{2\Omega_{\alpha0}}
 f_i^\alpha(t)f_j^{\alpha*}(t)
\simeq
\int\frac{d^3 k}{(2\pi)^3}\left\{\sum_\alpha \frac{1}{2\Omega_{\alpha0}}
f_{i0}^\alpha f_{j0}^{\alpha*}\right.
\\\nonumber
&&+\sum_{\alpha\beta}
\frac{1}{2\Omega_{\alpha 0}\Omega_{\beta 0}}
f_{i0}^\alpha f_{j0}^\beta\left[
-\frac{1}{\Omega_{\alpha0}+\Omega_{\beta0}}
\Bigl(\calv_{\alpha\beta}(t)-\calv_{\alpha\beta}(0)
\cos\left[(\Omega_{\alpha0}+\Omega_{\beta0})t\right]\Bigr)\right.
\\ \nonumber
&&+\frac{1}{(\Omega_{\alpha0}+\Omega_{\beta0})^2}
\dot\calv_{\alpha\beta}(0)
\sin\left[(\Omega_{\alpha0}+\Omega_{\beta0})t\right]
\\ \nonumber
&&+\frac{1}{(\Omega_{\alpha0}+\Omega_{\beta0})^3}
\left(\ddot\calv_{\alpha\beta}(t)-\ddot\calv_{\alpha\beta}(0)
\cos\left[(\Omega_{\alpha0}+\Omega_{\beta0})t\right]\right)
\\ 
&&\left.\left.
+\frac{1}{(\Omega_{\alpha0}+\Omega_{\beta0})^3}
\int dt' \stackrel{...}{\calv}_{\alpha\beta}(t')
\cos\left[(\Omega_{\alpha0}+\Omega_{\beta0})(t-t')\right]\right]\right\}\pkt
\eea
Several integrations by parts have been performed in order
to separate the high momentum power behavior.
The first term in the integrand is quadratically divergent, and 
the one proportional to $\calv(t)$ is logarithmically divergent.
In the process of renormalization these terms are removed 
and included in the mass and coupling constant renormalizations.
The contribution proportional to $\calv(0)$ vanishes as $\calv(0)=0$ by
definition, see Eq. \eqn{calvdef}. The terms proportional to $\dot\calv(t)$ and
$\ddot \calv(t)$ are finite at all times. The nonanalytic parts
are those proportional to
$\dot\calv(0)$ and $\ddot \calv(0)$. Near $t=0$ we find that the
momentum integrals which multiply $\dot\calv(0)$ and $\ddot \calv(0)$
behave as
\bea\nonumber
\int\frac{d^3k}{(2\pi)^3}
\frac{1}{2\Omega_{\alpha0}\Omega_{\beta0}(\Omega_{\alpha0}+
\Omega_{\beta0})^2} \sin[(\Omega_{\alpha0}+
\Omega_{\beta0})t]\simeq - \frac{1}{8 \pi^2}t \ln[(m_\alpha+m_\beta)t]\kma
\\\nonumber
\int\frac{d^3k}{(2\pi)^3}
\frac{1}{2\Omega_{\alpha0}\Omega_{\beta0}(\Omega_{\alpha0}+
\Omega_{\beta0})^3} \cos[(\Omega_{\alpha0}+
\Omega_{\beta0})t]\simeq \frac{1}{16\pi^2}t^2 \ln [(m_\alpha+m_\beta)t]
\pkt\eea
So in general the first and second derivatives of the 
fluctuation integrals  would be infinite at $t=0$.

As we have mentioned previously there are two methods for getting rid
of this singular behavior: either one introduces surface counterterms
or one modifies the initial state. Our approach is the second one,
and we have formulated this modification of the
initial state as a Bogoliubov transformation. The singular behavior
is obviously related to the large momentum behavior of the
integrand. So the modification of the initial state will constrain
only its ultraviolet behavior. We are still free to modify it
at finite momenta, or with contributions that vanish sufficiently fast
at large momenta, as e.g. a thermal initial state.

The Bogoliubov transformation and its consequences for the
fluctuation integral are presented in detail in  Appendix
\ref{bogotrans}. The general concept implies that we
replace our naive initial state, the vacuum state for quanta
of masses $m_{i0}$ by a transformed vacuum state, annihilated by
a superposition of annihilation operators $a_\alpha(\bfk)$ and
creation operators $a^\dagger_\alpha(-\bfk)$.
The essential formulae are:
\\
(i) the definition of the transformation
\be
\tilde a_\alpha(\bfk)=\sum_\beta
\sqrt{\frac{\Omega_{\alpha0}}{\Omega_{\beta0}}}
\left[ C^{\alpha\beta}a_\beta(\bfk)
- S^{\alpha\beta}a^\dagger_\beta(-\bfk)\right]
\, ;\ee
\\
(ii) the definition of a new vacuum state $|\tilde 0>$ via
\be
\tilde a _\gamma(\bfk)|\tilde 0> =
\sqrt{\frac{\Omega_{\gamma0}}{\Omega_{\alpha0}}}
 C^{\gamma\alpha}\left[a_\alpha(\bfk)-\sqrt{\frac{\Omega_{\alpha0}}
{\Omega_{\beta0}}}
\rho_{\alpha\beta}a_\beta^\dagger(-\bfk)\right]
|\tilde 0>=0
\kma\ee
where $ \rho= C^ {-1} S$ is a symmetric matrix;
\\
(iii) 
the definition of a matrix $M_{\alpha\beta}$ which is introduced
via the expectation value of $a_\alpha(\bfk)a^\dagger_\beta(\bfk')$
in the new vacuum:
\be
<\tilde 0|a_\alpha(\bfk)a^\dagger_\beta(\bfk')|\tilde 0>
=(2\pi)^3 \delta^3(\bfk-\bfk')2 \sqrt{\Omega_{\alpha0}\Omega_{\beta0}}
M_{\alpha\beta}(k)
\, ;\ee
and\\
(iv) the relation between $\rho$ and $M$ 
\be \label{rhoMrelation}
M - \rho M^{\sf T}  \rho^\dagger =I
\kma\ee
which ensures that the commutator of $\tilde a_\alpha(\bfk)$ and
$\tilde a^\dagger_\beta(\bfk')$ is canonical.
All matrices which we have introduced here  depend on $k=|\bfk|$.

As derived in Appendix \ref{bogotrans}, the fluctuation integral, 
when evaluated in the Bogoliubov-transformed vacuum, takes the form
\bea\nonumber
&&\tilde \calf_{ij}(t)=
<\tilde 0|\psi_i(\bfx,t)\psi_j(\bfx,t)|\tilde 0>
\\\nonumber
&=&\frac{1}{2}\int\frac{d^3 k}{(2\pi)^3}
\sum_{\alpha,\beta}
\frac{1}{2\sqrt{\Omega_{\alpha0}\Omega_{\beta0}}}
\left[f_i^\alpha(t)f_j^\beta(t)
\rho^{\beta\kappa}M_{\alpha\kappa}\right.
\\\nonumber
&&+f_i^\alpha(t)f_j^{\beta*}(t)
M_{\alpha\beta}
\\\nonumber
&&+f_i^{\alpha*}(t)f_j^{\beta}(t)
\rho^{\alpha\kappa*}\rho^{\beta\lambda}
M_{\kappa\lambda}
\\    \label{tildefdef}
&&\left.+ f_i^{\alpha*}(t)f_j^{\beta*}(t)
\rho^{\alpha\kappa*}
M_{\kappa\beta} \right]
\pkt\eea

We now have  to determine $\rho$  in such a way  as
to cancel the initial singularities which 
are contained in the integral over $f_i^\alpha(t)f_j^{\beta*}(t)$.
As discussed below Eq. \eqn{fluctint} the dangerous 
contributions are those involving \\
$\dot\calv_{\alpha\beta}(0)
\sin\left[(\Omega_{\alpha0}+\Omega_{\beta0})t\right]$ and
$\ddot\calv_{\alpha\beta}(0)
\cos\left[\Omega_{\alpha0}+\Omega_{\beta0})t\right]$. They have to be
cancelled by the terms proportional to $\rho$ and $\rho^2$
generated by the Bogoliubov transformation.
If one considers Eqs. \eqn{rhoMrelation} and \eqn{tildefdef}
one realizes that the determination
of $\rho$ seems to be  marred already by the nonlinear relation 
between $M$ and $\rho$. We have to realize, however, that there is
no unique choice for $\rho$, anyway. All we need is a
cancellation of the dangerous terms {\em at large momenta}.
These contributions are divided, in the integrand, by
combinations of $\Omega_{\alpha0}$ and $\Omega_{\beta0}$ which 
asymptotically behave as $k^{-4}$ and $k^{-5}$, respectively.
So these terms become small asymptotically, and to get
the correct asymptotic behavior of the matrix elements of $\rho$
we can work in the linear approximation.
In this approximation we have
$M_{\alpha\beta}\simeq\delta_{\alpha\beta}$,
$\rho^{\beta\kappa}M_{\alpha\kappa}\simeq\rho^{\beta\alpha}$,
and $M_{\kappa\lambda}\rho^{\alpha\kappa*}\rho^{\beta\lambda}\simeq 0$.
Furthermore, we can approximate $f_i^\alpha(t)\simeq f_{i0}^\alpha
\exp(-i\Omega_{\alpha0} t)$, whenever it appears multiplied by $\rho$.
Corrections would be of order
$\rho * \calv$; as $\calv$ is of order $\rho$ this
would be of order $\rho^2$.

When rewriting the fluctuation integral in the
Bogoliubov-transformed vacuum $\tilde\calf_{ij}$ we use the approximations
we have just mentioned.  We further use the expansion of 
 $f_i^\alpha(t)f_j^{\alpha*}(t)$ as it appears in Eq. \eqn{fluctint},
but we remove the renormalization parts.
We then obtain, to first order in $\calv$ and $\rho$
\bea\nonumber 
&&\tilde\calf_{ij}\simeq
\int\frac{d^3 k}{(2\pi)^3}\left\{
\sum_{\alpha,\beta} \frac{1}{2\sqrt{\Omega_{\alpha0}\Omega_{\beta0}}}
f_{i0}^\alpha f_{j0}^{\beta}\left[\rho^{\alpha\beta}
e^{-i(\Omega_{\alpha0}+\Omega_{\beta0})t}+
\rho^{\alpha\beta*}
e^{i(\Omega_{\alpha0}+\Omega_{\beta0})t}\right]\right.
\\\nonumber
&&
+\sum_{\alpha,\beta}
\frac{1}{2\Omega_{\alpha 0}\Omega_{\beta 0}}
f_{i0}^\alpha f_{j0}^\beta\left[
\frac{1}{(\Omega_{\alpha0}+\Omega_{\beta0})^2}
\dot\calv_{\alpha\beta}(0)
\sin\left[(\Omega_{\alpha0}+\Omega_{\beta0})t\right]\right.
\\ \nonumber
&&+\frac{1}{(\Omega_{\alpha0}+\Omega_{\beta0})^3}
\left(\ddot\calv_{\alpha\beta}(t)-\ddot\calv_{\alpha\beta}(0)
\cos\left[(\Omega_{\alpha0}+\Omega_{\beta0})t\right]\right)
\\ 
&&\left.\left.
+\frac{1}{(\Omega_{\alpha0}+\Omega_{\beta0})^3}
\int dt'\stackrel{...}{\calv}_{\alpha\beta}(t')
\cos\left[(\Omega_{\alpha0}+\Omega_{\beta0})(t-t')\right]\right]\right\}
\pkt\eea
The cancellation of the terms which would produce a singularity at $t=0$
then requires
\bea \label{rhoas1}
\Im \rho^{\alpha\beta}&=&\frac{1}{2\Omega_{\alpha 0}\Omega_{\beta 0}}
\frac{1}{(\Omega_{\alpha0}+\Omega_{\beta0})^2}
\dot\calv_{\alpha\beta}(0)\kma \\
\label{rhoas2}
\Re \rho^{\alpha\beta}&=&\frac{1}{2\Omega_{\alpha 0}\Omega_{\beta 0}}
\frac{1}{(\Omega_{\alpha0}+\Omega_{\beta0})^3}
\ddot\calv_{\alpha\beta}(0)
\pkt\eea
If the fields were independent $\calv_{\alpha\beta}$ would be
 diagonal and we would obtain
(omitting the indices of the diagonal elements)
\be
\frac{\Im \rho}{\Re \rho}=2 \Omega_0 \frac{\dot\calv(0)}{\ddot\calv(0)}
\ee
and 
\be
|\rho|=\frac{1}{8\Omega_0^3}\sqrt{\dot\calv(0)^2+\frac{\ddot\calv(0)^2}{4
\Omega_0^2} }
\ee
for the separate Bogoliubov transformations of the two fields.
This agrees in the approximation of large momenta with
the results for the one-field case,
Eqs. (51) and (52) of Ref. \cite{Baacke:1997zz}. There it 
was possible to remove the contributions proportional to $\dot \calv(0)$
and $\ddot \calv(0)$ for all momenta. Here these terms are cancelled
at large momenta only. 

With  Eqs. \eqn{rhoas1} and \eqn{rhoas2} we have obtained 
a solution to our problem of initial singularities. We have to stress
that there is an infinite manifold of such solutions, differing, e.g.,
by a different choice of initial occupation numbers at finite momenta. 
They all have to share the same large momentum behavior, however.

Once we have  $\rho$ we now must determine $M$, using \eqn{rhoMrelation},
 {\em without any approximation}, because otherwise
our transformation would not be canonical. Though $\rho$ appears
nonlinearly, Eq.\eqn{rhoMrelation} is simply a system of four linear equations
for the matrix elements of $M$. Finally, the
fluctuation integral \eqn{tildefdef} has to be computed using the
exact numerical solutions $f_i^\alpha(t)$ in all four
terms of the integrand.

For the numerical computations it is preferable to
implement the Bogoliubov transformation in a different way, by redefining the
mode functions. For this purpose we introduce
\be\label{ftildedef}
\tilde f^\gamma_i(k,t)=\sum_\alpha\frac{\sqrt{2\Omega_{\gamma0}}}
{\sqrt{2\Omega_{\alpha0}}}
\left[ f_i^\alpha(k,t)N_{\alpha\gamma} 
+f^{\alpha *}_i(k,t)\rho^{\alpha\kappa*}N_{\kappa \gamma}\right]
\kma\ee
where the matrix $N$ satisfies $ N\times N =M$
\footnote{$N$ is not uniquely
determined, all we need is one particular matrix that 
satisfies this relation. As $M$ is Hermitian, so is $N$.}.
It can be determined using the eigenvalues and eigenvectors of $M$.
One easily verifies that 
\be \label{tildefdef2} 
\tilde \calf_{ij}(t)=\sum_\alpha
\int\frac{d^3k}{(2\pi)^3 2\Omega_{\gamma0}
}
\tilde f_i^\gamma(k,t)\tilde f_j^{\gamma*}(k,t) 
\ee
is identical to the previous definition, Eq. \eqn{tildefdef}.
As $ f_i^\alpha $ and $ f_i^{\alpha*}$ are solutions of the
same equation of motion, Eq. \eqn{fdgl}, so is $\tilde f_i^\gamma$. It is
sufficient, therefore, to determine $\tilde f_i^\gamma$ 
as a solution to this equation with the initial conditions 
implied by Eqs. \eqn{ftildedef},\eqn{finit} and
\eqn{fpinit}.
 
Having presented the technical procedure we would like to add some
comments  concerning the interpretation. In order to do so we need 
to discuss the problem of initial states in a more general way.

The adiabatic vacuum has often been used as an initial state for
preheating simulations, maybe on the grounds that after inflation one ends
up in a temperature zero state, i.e. an ``empty'' vacuum. 
Indeed if this is the case, and if the
evolution of the classical fields is very slow (``adiabatic''), 
this can be considered to be a reasonable
guess for an initial state. Another choice that may be reasonable,
e.g. after thermalization and in a period of adiabatic evolution, 
is a thermal initial state. This is of course not a pure state.
For a thermal state the fluctuation integral would be replaced by
\be \label{particlestate}
\calf_{ij}(t)
=\sum_\alpha
\int\frac{d^3k}{(2\pi)^3 2 \Omega_{\alpha0}(k)}(2 N_\alpha(k)+1)
f_i^\alpha(k,t)f_j^{\alpha*}(k,t)
\kma
\ee
with
\be
N_\alpha(k)=\left[\exp(\Omega_\alpha(k)/T)-1\right]^{-1}
\pkt
\ee
Here we have used the fact that the mass matrix and therefore 
the fluctuation Hamiltonian is diagonal in the basis $f_{i0}^\alpha$ 
at $t=0$. 

If one takes into account the real evolution of the system before
$t=0$ then neither the adiabatic vacuum nor a thermal initial state 
will be appropriate. If the system has started, before $t=0$,
in a pure quantum state, it cannot have ended up, at
$t=0$, in a thermal state
or in any other state described simply by particle numbers
$N_\alpha(k)$. The quantum system can be interpreted as a system
of independent free particle only after ``decoherence'', a concept
that has been addressed in the present context in
Refs. \cite{Polarski:1995jg,Khlebnikov:1996mc}. But even if 
the system has started, at an earlier time,
with a mixed state, the interaction with the background field will
have created a coherence in the different components of such a state
at $t=0$,  and a representation in the form \eqn{particlestate}
 will not be possible.

So, if one takes into account the evolution of the system prior to
$t=0$ then one would have to  know the entire prehistory or at least the
prehistory of a long period in order to describe the state at
$t=0$ with its full quantum coherence. This is of course not 
possible unless one knows how to start the system at an earlier time,
facing then the same problem. The best one can hope for
is that after some time the system will not remember much of its initial state.
This is presumably the case if the background fields produce
large quantum fluctuations at later times.

The purpose of the Bogoliubov transformation is different.
If we know the evolution of the background fields near $t=0$ 
(and by continuity this means also shortly before $t=0$), to the extent 
that we know $\dot \calv(0)$ and $\ddot \calv(0)$, or, equivalently,
$\dot \phi(0)$ and $\ddot \phi(0)$,  then we have a limited 
information on the initial state. Constant background fields at
$t < 0$ and the adiabatic vacuum state as initial state would 
produce a singularity of the first two derivatives of the Green's function.
The Bogoliubov transformation removes this singularity, or, more precisely,
it reduces it to higher orders in the derivatives. In this way it takes
a minimal account of the fact that the system is not static before
$t=0$. As we have
displayed above the transformation
may be considered as a modification of the state or of the mode functions.
The new state should not be considered as a vacuum state. The adiabatic vacuum 
state remains the lowest energy state for a given set of background fields.
The analysis of the high momentum behavior of the fluctuation integral
simply shows that the system will never arrive at this state if the
background fields keep changing with time. In reality, of course,
 we would rather expect
the quantum state of the fluctuations to be an excited one, particularly
at low momentum. Our simple requirement of continuity for the
Green's function does not give us any information on this excited
state, {\em except at high momenta}.

The state generated by the Bogoliubov transformation applied to the
adiabatic vacuum is a pure state. Therefore, it  {\em cannot} 
be described by a mixed state with suitable particle numbers $N_\alpha(k)$.
If for some physical motivation we want to
start with a thermal state or some other state specified by particle
numbers we have to combine two different concepts: a mixed state
made up of different excited Fock-space states and a pure state
that ensures the continuity of the Green' s function. For a thermal state
the particle numbers decrease exponentially as $k \to \infty$;
then the discontinuity of the Green' s function solely arises
from the vacuum contribution. Its Bogoliubov transformation is
well-defined and compulsory at high momenta only. 
There are then, among many others, two pragmatic ways of 
defining a thermal initial state:
(i) One defines the thermal state using for all momenta
the modified mode functions $\tilde f_i^\alpha(k,t)$. This is 
not a quite a thermal state, though, as
the modified mode functions are not eigenfunctions of 
$i \partial/\partial t$. (ii) One uses the integrand
of Eq. \eqn{particlestate} with the original mode
functions $f_i^\alpha(k,t)$ for low momenta only, and 
the integrand of Eq. \eqn{tildefdef2} at high momenta.


\section{The energy-momentum tensor}
\label{energymomentumtensor}
\setcounter{equation}{0}
The energy-momentum tensor for the fluctuations
$t_{\mu\nu}=(T^{\rm fluc})_{\mu\nu}$ in a homogeneous
background field is diagonal and has identical space-space
components.
It may be specified by the two independent expectation values
\bea\label{t00def}
t_{00}= \epsilon&=&\frac12 <
\dot \psi_i(\bfx,t)\dot \psi_i(\bfx,t)\\
\nonumber&& +
\vec \nabla \psi_i(\bfx,t)\vec\nabla \psi_i(\bfx,t)+
 \calm^2_{ij}(t)\psi_i(\bfx,t)\psi_j(\bfx,t)>
\eea
and
\bea\label{tmumudef}
t_\mu^\mu=\epsilon-3p&=&
<-\dot \psi_i(\bfx,t)\dot \psi_i(\bfx,t)\\
\nonumber&& +
\vec \nabla \psi_i(\bfx,t)\vec\nabla \psi_i(\bfx,t)+
 2\calm^2_{ij}(t)\psi_i(\bfx,t)\psi_j(\bfx,t)>
\pkt\eea
$\epsilon$ is the energy density and $p$ the pressure.
$t_{00}$ and $t_\mu^\mu$ 
can be evaluated in the adiabatic vacuum and in the Bogoliubov-transformed
vacuum in the same way as the fluctuation integrals. 
We just present the expectation values in the
Bogoliubov-transformed vacuum, the one in the adiabatic vacuum is
obtained by substituting $\rho\to 0$ and $M\to I$.
For $\tilde t_{00}$ we obtain
\bea\nonumber
&&\tilde t_{00}=\int\frac{d^3k}{(2\pi)^3}
\sum_{\alpha, \beta}\frac{1}{4\Omega_{\alpha0}}\left\{
\Re \left[\rho^{\beta\kappa}M_{\alpha\kappa}
\left(\dot f_i^\alpha\dot f_i^\beta+
k^2f_i^\alpha f_i^\beta+\calm^2_{ij}f_i^\alpha f_j^\beta\right)\right]
\right.\\\label{t00tildedef}
&&\left. +\left[M_{\alpha\beta}+M_{\kappa\lambda}\rho^{\beta\kappa*}
\rho^{\alpha\lambda}\right]
\left(\dot f_i^\alpha\dot f_i^{\beta*}+
k^2f_i^\alpha f_i^{\beta*}+\calm^2_{ij}f_i^\alpha f_j^{\beta*}\right)
\right\}
\pkt\eea
For the trace we find
\bea\nonumber
&&\tilde t_\mu^\mu=\int\frac{d^3k}{(2\pi)^3}
\sum_{\alpha,\beta}\frac{1}{4\Omega_{\alpha0}}\left\{
\Re \left[\rho^{\beta\kappa}M_{\alpha\kappa}
\left(-\dot f_i^\alpha\dot f_i^\beta+
k^2f_i^\alpha f_i^\beta+2\calm^2_{ij}f_i^\alpha f_j^\beta\right)\right]
\right.\\\label{tmumutildedef}
&&\left. +\left[M_{\alpha\beta}+M_{\kappa\lambda}\rho^{\beta\kappa*}
\rho^{\alpha\lambda}\right]
\left(-\dot f_i^\alpha\dot f_i^{\beta*}+
k^2f_i^\alpha f_i^{\beta*}+2\calm^2_{ij}f_i^\alpha f_j^{\beta*}\right)
\right\}
\pkt\eea
Both expressions can alternatively be rewritten in terms of the modified mode
functions of Eq. \eqn{ftildedef}.

Using the equation of motion for the fluctuations we can write
\bea
-\dot f_i^\alpha\dot f_i^{\beta*}+
k^2f_i^\alpha f_i^{\beta*}+\calm^2_{ij}f_i^\alpha f_j^{\beta*}&=&
-\frac12 \frac{d^2}{dt^2}f_i^\alpha f_j^{\beta*}\kma
\\
-\dot f_i^\alpha\dot f_i^{\beta}+
k^2f_i^\alpha f_i^{\beta}+\calm^2_{ij}f_i^\alpha f_j^{\beta}&=&
-\frac12 \frac{d^2}{dt^2}f_i^\alpha f_j^{\beta}
\pkt\eea
Therefore the trace can be expressed in terms of the fluctuation
integrals \eqn{tildefdef} as 
\be\label{tmumutilde}
\tilde t_\mu^\mu
=-\frac{1}{2}\frac{d^2}{dt^2}\tilde \calf_{ii}+\calm_{ij}\tilde
\calf_{ij}
\pkt\ee
Both the energy density and the trace contain second space and time
derivatives of the 
two-point function and this can transform the mild singularities 
found in the fluctuation integrals $\calf_{ij}$ into infinities at $t=0$.
In Ref. \cite{Baacke:1997zz} it was found that the energy density
remains finite even in the adiabatic vacuum.
As the second derivatives only appear in the kinetic terms which are
diagonal this analysis remains valid for the coupled-channel case.
However, the trace of the energy-momentum tensor $t^\mu_\mu$ in
the adiabatic vacuum contains the second time derivative of the
fluctuation integrals $\calf_{ii}$ and this behaves
as $d^2 (t \ln t)/dt^2=1/t$ as $t\searrow 0$.
In the transformed fluctuation integrals $\tilde \calf_{ij}$ we have removed
the dangerous terms, and  so $\tilde t^\mu_\mu$ has a finite value at $t=0$.
 
When the field is coupled to gravity \cite{Baacke:2009?} the expressions 
\eqn{t00def} and \eqn{tmumudef} receive some further contributions that
 we do not want to discuss here in detail. They  can be written in terms
of the fluctuation integrals and their first derivatives.
The fluctuation integrals themselves are  not infinite at $t=0$. 
The most singular of the additional terms
are proportional to  $(\xi_i-1/6)H d\calf_{ii}/dt$ and appear both
in $t_{00}$ and $t^\mu_\mu$. They behave as $\ln t$ as $t\searrow 0$; in
 the Bogoliubov-transformed initial state $\calf_{ii}$ is
replaced by $\tilde \calf_{ii}$ and then the energy-momentum tensor
remains finite at $t=0$.


\section{The adiabatic particle number}
\label{particles}
\setcounter{equation}{0}
The  adiabatic particle number is obtained by representing the
fluctuation field  at time $t$ in terms of the adiabatic
Fock space at time $t$.
The fluctuation field is 
given by Eq. \eqn{fieldexp0}. The adiabatic Fock 
space at time $t$ is defined in terms of particle excitations
which are eigenstates of the mass matrix $\calm^2_{ij}(t)$.
We define the eigenvectors of the mass matrix by
\be
\calm^2_{ij}(t)f_{jt}^\alpha=m^2_{\alpha t}f^\alpha_{it}
\kma\ee
we again choose them to be real and normalized via
\be\label{normt}
\sum_i f^\alpha_{i t}f^\beta_{i t}=\delta^{\alpha\beta}
\ee
and define $\Omega_{\alpha t}=\sqrt{k^2+m^2_{\alpha t}}$.
We further expand the fields with respect to the new basis
as
\bea\label{fieldexpt}
\psi_i(\bfx,t)&=&\sum_\alpha
\int \frac{d^3 k}{(2\pi)^32\Omega_{\alpha t}}
\left[a_{\alpha t}(\bfk)f^\alpha_{ i t}
+a^\dagger_{\alpha t}(-\bfk)f^\alpha_{i t}\right]e^{i\bfk\bfx}\kma
\\
\dot\psi_i(\bfx,t)&=&-i \sum_\alpha
\int \frac{d^3 k}{(2\pi)^3 2 }
\left[a_{\alpha t}(\bfk)f^\alpha_{i t}
-a^\dagger_{\alpha t}(-\bfk)f^\alpha_{i t}\right]
e^{i\bfk\bfx}
\kma\eea
where we have chosen the initial conditions for the modes $f^\alpha_{it}(k,t)$
in analogy to Eqs. \eqn{finit} and \eqn{fpinit}.
Using the field expansion the new annihilation  
operators $a_{\alpha t}(\bfk)$
can be expressed as
\be \label{atdef}
a_{\alpha t}(\bfk)=\int d^3 x e^{-i\bfk\bfx}
\left[\Omega_{\alpha t}\psi_i(\bfx,t)
+i\dot \psi_i(\bfx,t)\right]f_{i t}^\alpha
\pkt\ee 
They annihilate the adiabatic vacuum defined at time $t$.
The relation to the original operators $a_\alpha(\bfk)$ is obtained
by inserting the field expansion \eqn{fieldexp0} into Eq. \eqn{atdef}.
We find
\be
 a_{\alpha t}(\bfk)=\sum_\beta
\sqrt{\frac{\Omega_{\alpha t}}{\Omega_{\beta 0}}}
\left[C^{\alpha\beta}a_{\beta }(\bfk)
-S^{\alpha\beta}a^\dagger_{\beta }(-\bfk)\right]
\kma\ee
with
\bea\label{Ct}
C^{\alpha\beta}&=&
\frac{1}{2\sqrt{\Omega_{\alpha t}\Omega_{\beta0}}}
\left[\Omega_{\alpha t} f_i^\beta(k,t)+i\dot f^\beta_i(k,t)\right]
f_{i t}^\alpha \kma
\\\label{St}
S^{\alpha\beta}&=&
\frac{-1}{2\sqrt{\Omega_{\alpha t}\Omega_{\beta0}}}
\left[\Omega_{\alpha t} f_i^{\beta*}(k,t)+i\dot f^{\beta*}_i(k,t)\right]
f_{i t}^\alpha
\pkt\eea
Using the relations \eqn{ff}, \eqn{ffp} and \eqn{fpfp} of Appendix
\ref{canform} it is straightforward to
verify that this is a Bogoliubov transform, i.e., that
Eqs. \eqn{CSSC} and \eqn{CCSS} are satisfied.
In terms of the matrices $C$ and $S$
the adiabatic particle number density is given by \footnote{No summation over
$\alpha$. For simplicity of presentation we assume 
the adiabatic vacuum as initial state.},
\be \label{nalphadef}
n_\alpha(k,t)=\frac{1}{2 V \Omega_{\alpha t}}
<0|a^\dagger_{\alpha t}(\bfk)a_{\alpha t}(\bfk)|0>
=\sum_\gamma
S^{\alpha\gamma*}S^{\alpha\gamma}
\pkt\ee
Inserting Eq.  \eqn{St} we obtain
\be
n_\alpha(k,t)=\frac{1}{2\Omega_{\alpha t}}
\sum_\gamma \frac{1}{2\Omega_{\gamma 0}}
\left[\Omega_{\alpha t}^ 2f_i^{\gamma}f_j^{\gamma*}+
\dot f_i^{\gamma}\dot f_j^{\gamma*}\right]f_{it}^\alpha f_{jt}^\alpha
\pkt
\ee
This has a simple interpretation: one decomposes the energy density
with respect to the fluctuations $f^\alpha_{it}$. Then $n_\alpha(k,t)$
is obtained by dividing the part corresponding to the
fluctuations $f^\alpha_{it}$ by the frequency $\Omega_{\alpha t}$
of these fluctuations. This result is analogous to the one-channel case.

While the definition \eqn{nalphadef} is suggestive we would like
to add that this particle number does not imply a representation
of the fluctuation integral in the form \eqn{particlestate}
in terms of the mode functions $f^\alpha_{it}(k,t)$.
Indeed, if one wants to use the representation
\eqn{fieldexpt} for calculating the fluctuation integral one gets
nontrivial contributions from the operators
 $a_{\alpha t}(\bfk)a_{\alpha t}(\bfk')$,
$a^\dagger_{\alpha t}(\bfk)a^\dagger_{\alpha t}(\bfk')$ and
$a^\dagger_{\alpha t}(\bfk)a^\dagger_{\alpha t}(\bfk')$ 
as well, see Eqs. \eqn{aaexval},
\eqn{adaggeradaggerexval} and \eqn{adaggeraexval}.
These contributions to the fluctuation integral
 are negligible for large particle numbers only, i.e., if the
matrix elements $S^{\alpha\gamma}$ are much larger than the 
$C^{\alpha\gamma}$. 


\section{Summary}
\setcounter{equation}{0}
\label{summary}
We have addressed here two topics of the nonequilibrium dynamics
of coupled fields in a one-loop approximation
to quantum backreaction: the problem of the
initial time singularity in the energy-momentum tensor
and the definition of the
adiabatic particle number for a  system of coupled
scalar fields. Along with these topics  we have considered 
Bogoliubov transformations
and some aspects of the canonical formalism for such coupled
systems.
 
Our main interest, as evident from the title, were the initial
time singularities. We have been able to define a Bogoliubov
transformation of the initial state that removes the initial time
singularities in such a way that the energy-momentum tensor
is finite in the limit $t\searrow 0$. This is important if one
considers the evolution of such a system of fields coupled to
gravity. Clearly, this Bogoliubov transformation 
is constrained only at large momenta. So modifications
that are subleading at high momenta are still acceptable.
We had to include a discussion of the canonical formalism for
a coupled-channel system, as some of the results were needed in
the construction  of the initial time singularities: we had to ensure
that the fluctuation integrals are real and symmetric in the indices,
as they should be on account of their definition. 

Both the discussion
of the canonical formalism and of Bogoliubov transformations
for coupled-channel systems are at the same time the basis for defining
the adiabatic particle number density. So we have derived an expression for
this density in terms of the coupled system mode functions. It is analogous to
the definition in the single-channel case and has a simple intuitive interpretation.
 Another formulation for the adiabatic particle number, based in an eikonal formalism
and the evolution of Bogoliubov coefficients,
has been presented recently \cite{Nilles:2001fg}. As both formalisms are
canonical, the results should be equivalent, though it may be difficult
to verify this analytically.


\section*{Acknowledgments}
One of us (N.K.) thanks the Humboldt Foundation for financial
support, and the Deutsche Elektronensynchrotron DESY, Hamburg,
 for hospitality.


\begin{appendix}

\section{The Bogoliubov transformation for a coupled system}
\label{bogotrans}
\setcounter{equation}{0}
We first recall some basic relations for the case of a single
quantum field, see, e.g., Ref. \cite{Bogoliubov:1983}.
The Bogoliubov transformation rotates creation into annihilation
operators and vice versa, such as to preserve the canonical
commutation relations
\bea \nonumber
\left[a(\bfk),a(\bfk')\right] &=&0\kma \,
\\
\left[a(\bfk),a^\dagger(\bfk')\right]&=&(2\pi)^3 2\omega \delta^3(\bfk-\bfk')\kma
\,
\\
\nonumber
\left[a^\dagger(\bfk),a^\dagger(\bfk')\right]&=&0
\pkt\eea
 Furthermore the transformation has to be
chosen in such a way that the vacuum retains its
total momentum zero and remains isotropic.
The most general form of such a transformation is then
induced by the operator
\be
Q=\frac{1}{2}\int\frac{d^3k}{(2 \pi)^3 2\omega}
\left[q(k) a^\dagger(\bfk)a^\dagger(-\bfk)
-q^*(k)  a(\bfk)a(-\bfk)\right]\kma
\ee
via
\be
\tilde a(\bfk)=\exp(Q)a(\bfk)\exp(Q^\dagger)=\exp(Q)a(\bfk)\exp(-Q)
\ee
and
\be
|\tilde 0> =\exp (Q) |0>
\pkt\ee
Here $q(k)$ is a general complex function of $k=|\bfk|$.We have
\bea
\left[ a(\bfk),Q\right] &=&q(k)a^\dagger(-\bfk)\kma
\\
\left[ a^\dagger(-\bfk),Q\right] &=&q^*(k)a(\bfk)
\pkt
\eea
We have in general
\be 
a(\bfk)\exp(-Q)=
\exp(-Q)\sum_{n=1}^\infty\frac{(-1)^n}{n!}[[[[[a(\bfk),Q],Q]...],Q]_n
\kma\ee
where the $n$-th term in the sum contains $n$ commutators.
The even commutators $(n=2l)$ yield
\be
[[[[[a(\bfk),Q],Q]...],Q]_{2l}=|q(k)|^{2l}a(\bfk)
\kma\ee
the odd ones $(n=2l+1)$ yield
\be
[[[[[a(\bfk),Q],Q]...],Q]_{2l+1}=|q(k)|^{2l}q(k) a^\dagger(-\bfk)
\pkt\ee
Writing $q(k)=\gamma(k)e^{i\delta(k)}$ with real
constants $\gamma$ and $\delta$
we find
\be
a(\bfk)\exp(-Q)=\exp(-Q)\left[\cosh(\gamma)a(\bfk)
-\sinh(\gamma) e^{i\delta}a^\dagger(-\bfk)\right]
=\exp(-Q)\tilde a(\bfk)
\pkt\ee

With these preliminaries the generalization is straightforward.
We have two sets of creation and annihilation
operators $a^\dagger_\alpha(\bfk)$ and
$a_\alpha(\bfk)$, where $\alpha=1,2$ labels the
two independent solutions $f_i^\alpha(k)$.
We have the field expansion
\be
\psi_i(\bfx,t)=
\int\frac{d^3 k}{(2\pi)^3}\sum_\alpha \frac{1}{2 \Omega_{\alpha0}}
\left\{a_\alpha(\bfk)f_i^\alpha(k,t)e^{i\bfk\bfx}
+a^\dagger_\alpha(-\bfk) f_i^{\alpha*}(k,t)e^{-i\bfk\bfx}\right\}
\ee
and the commutation relations
\be
\left[ a_\alpha(\bfk),a^\dagger_\beta(\bfk')\right]
=(2\pi)^3 2 \Omega_{\alpha0} \delta_{\alpha\beta}\delta^3(\bfk-\bfk')
\pkt\ee
The operator $Q$ now takes the form
\be
Q=\frac{1}{2}\int\frac{d^3k}{(2 \pi)^3}
\sum_{\alpha,\beta}\frac{1}{ 2\sqrt{\Omega_{\alpha0}\Omega_{\beta0}}}
\left[q^{\alpha\beta}(k) a_\alpha^\dagger(\bfk)a_\beta^\dagger(-\bfk)
-q^{\alpha\beta *}(k)  a_\alpha(\bfk)a_\beta(-\bfk)\right]
\pkt\ee
The normalization convention introduced by writing
$\sqrt{\Omega_{\alpha0}\Omega_{\beta0}}$ has the advantage of keeping
the functions $q^{\alpha\beta}(k)$ symmetric in the indices.
Indeed this symmetry is the only restriction on these
functions; as they are complex we have {\em six free
parameters}, which are functions of $k$. 
The symmetry arises from the fact that 
the products $a^\alpha(\bfk)a^\beta(-\bfk)$ and 
 $a^\beta(\bfk)a^\alpha(-\bfk)$ are equivalent. On the one hand
the operators commute, and on the other hand the arguments
$\bfk$ and $-\bfk$ may be exchanged as the integration
is symmetric in the sign of $\bfk$ and the functions
$q^{\alpha\beta}$ only depend on $|\bfk|$. An asymmetric
part of these functions would simply be summed and integrated
away. We again have $Q^\dagger=-Q$ and the transformation
matrix $\exp(Q)$ is unitary.

What does not work here, at least not in a general
parameterization $q^{\alpha\beta}$, is the 
explicit evaluation of the transformation of the
annihilation and creation operators.
The matrix $q^{\alpha\gamma}q^{\gamma\beta *}$ which appears 
after every second step in the evaluation of the 
multiple commutators, is given by
\be
q^{\alpha\gamma}q^{\gamma\beta *}
=\left(
\begin{array}{cc}
\left|q^{11}\right|^2+ \left|q^{12}\right|^2 &
q^{11}q^{12*}+q^{12}q^{22*}\\
q^{11*}q^{12}+q^{12*}q^{22}&   \left|q^{12}\right|^2+ \left|q^{22}\right|^2
\end{array}\right)^{\alpha\beta}
\pkt\ee 
It is Hermitian, in analogy to the reality of $|q|^2$ in the
single-channel case. It is diagonal
 in two cases: (i) $q^{12}=0$ and (ii) $q^{11}=q^{22}=0$.
It is instructive to evaluate the transformation of 
$a_\alpha(\bfk)$ in the two cases.
In the first case we find
\bea
\tilde a_1(\bfk)&=&\cosh(|q^{11}|) a_1(\bfk)-\sinh(|q^{11}|)
e^{i\arg(q^{11})}a_1^\dagger(-\bfk)
\kma\\
\tilde a_2(\bfk)&=&\cosh(|q^{22}|) a_2(\bfk)-\sinh(|q^{22}|)
e^{i\arg(q^{22})}a_2^\dagger(-\bfk)
\kma\eea
i.e., a simple Bogoliubov transformation for each
channel.
In the second case we have
\bea
\tilde a_1(\bfk)&=&\cosh(|q^{12}|) a_1(\bfk)-\sinh(|q^{12}|)
e^{i\arg(q^{12})}\sqrt{\frac{\Omega_{10}}{\Omega_{20}}}
a_2^\dagger(-\bfk)
\kma\\
\tilde a_2(\bfk)&=&\cosh(|q^{12}|) a_2(\bfk)-\sinh(|q^{12}|)
e^{i\arg(q^{12})}\sqrt{\frac{\Omega_{20}}{\Omega_{10}}}
a_1^\dagger(-\bfk)
\kma\eea
i.e., an annihilation operator in channel $1$ is mixed with
a creation operator in channel $2$.

In the general case the matrix $q^{\alpha\gamma}q^{\gamma\beta *}$
is not diagonal. Still we can sum up the series formally,
as a series of matrix products. As the exponential series
converges well this can be done even numerically.
We write
\be
\tilde a_\alpha(\bfk)=\sum_\beta
\sqrt{\frac{\Omega_{\alpha0}}{\Omega_{\beta0}}}
\left[C^{\alpha\beta}a_\beta(\bfk)
-S^{\alpha\beta}a^\dagger_\beta(-\bfk)\right]
\pkt\ee
In terms of the matrix $q^{\alpha\beta}$
we then have
\be 
C^{\alpha\beta}=\sum_n \frac{1}{(2n)!}\left[(qq*)^n\right]^{\alpha\beta}
\kma\ee 
where $qq*$ is the matrix product $q^{\alpha\gamma}q^{\gamma\beta *}$ and the 
power series is a  series of powers of this matrix.
Further, we have
\be 
S^{\alpha\beta}=\sum_n \frac{1}{(2n+1)!}\left[(qq*)^n\right]^{\alpha\gamma}
q^{\gamma\beta}
\pkt
\ee
Instead of writing these matrices as power series in $q^{\alpha\beta}$
we can ask for the conditions on $C$ and $S$ that follow from the
requirement that the commutation rules should be conserved.
From
\be
\left[a_\alpha(\bfk),a_\beta(\bfk')\right]=0
\ee
one finds
\be\label{CSSC}
\sum_\gamma\left(C^{\alpha\gamma}S^{\beta\gamma}
 -S^{\alpha\gamma}C^{\beta\gamma}\right)
=0
\kma \ee 
or
\be\label{CSSCmatrix}
C S^{\sf T}=S C^{\sf T}=\left (C S^{\sf T}\right)
^{\sf T}
\pkt\ee
Multiplying from the left with $ C^{-1}$ and from the right with
$( C^{\sf T})^{-1}$ one finds
\be \label{rhohatreal}
C^{-1}S=\left( C^{-1}S\right)^{\sf T}
\kma \ee
i.e., this a symmetric matrix.

Considering the nonvanishing commutator we find
\be\label{CCSS}
\sum_\gamma
\left(C^{\alpha\gamma} C^{\beta \gamma *}-
S^{\alpha\gamma} S^{\beta \gamma *}\right)=
\delta_{\alpha\beta}
\kma \ee
or, in matrix form,
\be\label{CCSSmatrix}
C C^\dagger-S  S^\dagger= I
\kma\ee
the obvious generalization of
\be
\cosh^2 (\gamma) - \sinh^2 (\gamma)=1
\pkt\ee
Instead of having to deal with two matrices it may be more convenient to deal
with just one: the condition that the operators $\tilde a_\alpha (\bfk)$
annihilate the vacuum $|\tilde 0>$ reads
\be \label{conditionwithrho}
\tilde a_\gamma(\bfk)|\tilde 0> =
\sqrt{\frac{\Omega_{\gamma0}}{\Omega_{\alpha0}}}C^{\gamma\alpha}\left[
  a_\alpha(\bfk)-\sqrt{\frac{\Omega_{\alpha0}}{\Omega_{\beta0}}}
\rho^{\alpha\beta}a^\dagger_\beta(-\bfk)\right]
|\tilde 0>=0
\kma \ee 
where we have defined the matrix
\be
\rho=C^{-1}S
\pkt \ee
From Eq. \eqn{rhohatreal} we see that
$\rho$ is a symmetric matrix. Indeed we had  found
previously that we have six free parameters for the most general
transformation.

To begin with we compute the expectation value of
$a_\alpha(\bfk)a_\alpha^\dagger(\bfk')$ in the new vacuum.
As vacua are homogeneous and isotropic we can write
\be \label{aadaggerexval}
<\tilde 0|a_\alpha(\bfk)a^\dagger_\beta(\bfk')|\tilde 0>
=(2\pi)^3 \delta^3(\bfk-\bfk')2 \sqrt{\Omega_{\alpha0}\Omega_{\beta0}}
M_{\alpha\beta}(k)
\pkt\ee
This definition implies that $M$ is a Hermitian matrix.
Using the commutation relations and \eqn{conditionwithrho}
we have
\bea
&&<\tilde 0|a_\alpha(\bfk)a^\dagger_\beta(\bfk')|\tilde 0>
\\\nonumber &=&
(2\pi)^3 2\Omega_{\alpha0}\delta^3(\bfk-\bfk')\delta_{\alpha\beta}+
<\tilde 0|a_\beta^\dagger(\bfk')a_\alpha(\bfk)|\tilde 0>
\\\nonumber
&=&
(2\pi)^32\Omega_{\alpha0} \delta^3(\bfk-\bfk')\delta_{\alpha\beta}
+\sqrt{\frac{\Omega_{\alpha0}\Omega_{\beta0}}
{\Omega_{\kappa0}\Omega_{\lambda0}}}\rho^{\alpha\kappa}\rho^{\beta\lambda*}
<\tilde 0|a_\lambda(-\bfk')a^\dagger_\kappa(-\bfk)|\tilde 0>
\pkt\eea
In terms of the matrix $M$ we find
\bea
M_{\alpha\beta}\sqrt{\Omega_{\alpha0}\Omega_{\beta0}}
=
\delta_{\alpha\beta} \Omega_{\alpha0}+\sqrt{\Omega_{\alpha0}\Omega_{\beta0}}
\rho^{\alpha\kappa}\rho^{\beta\lambda*}
M_{\lambda\kappa}
\kma \eea
or
\be\label{Meq}
M -\rho M^{\sf T}\rho^\dagger =I
\pkt\ee
This can be  solved  explicitly for $M$. 
Using the symmetry of $\rho$ it easy to
verify, e.g., using the series expansion in $\rho\rho^\dagger$, that
\be
M = (I-\rho\rho^\dagger)^{-1}
\pkt\ee
Furthermore, using $S=C\rho$ it is easy to see, using Eq. \eqn{CCSSmatrix},
 that
\be
CC^\dagger= (I-\rho\rho^\dagger)^{-1}=M
\pkt
\ee
This implies that knowing $\rho$ the matrix $C$ is not determined uniquely.
If one uses a basis in which $\rho$ is diagonal, $CC^\dagger$ is diagonal
 as well, and we have two free phases in the
matrix $C$. A further useful identity is
\be \label{MrhorhoMT}
M\rho=\rho M^{\sf T}
\pkt\ee 
It can easily be verified using again the expansion of $M$ interms
of $\rho\rho^\dagger$.

We next evaluate the expectation values of the other products:
\bea\nonumber
&&<\tilde 0|a_\alpha(\bfk)a_\beta(\bfk')|\tilde 0>=
\sqrt{\frac{\Omega_{\beta0}}{\Omega_{\kappa0}}}
\rho^{\beta\kappa} <\tilde 0|a_\alpha(\bfk)a^\dagger_\kappa(-\bfk')|\tilde 0>
\\ 
&=& \label{aaexval}(2\pi)^3\delta^3(\bfk+\bfk')
2\sqrt{\Omega_{\alpha0}\Omega_{\beta0}}
\rho^{\beta\kappa}M_{\alpha\kappa}(k)
\kma\eea
\bea\nonumber
&&<\tilde 0|a^\dagger_\alpha(\bfk)a^\dagger_\beta(\bfk')|\tilde 0>=
\sqrt{\frac{\Omega_{\alpha0}}{\Omega_{\kappa0}}}\rho^{\alpha\kappa*} <\tilde 0|a_\kappa(-\bfk)a^\dagger_\beta(\bfk')|\tilde 0>
\\ \label{adaggeradaggerexval}
&=&(2\pi)^3\delta^3(\bfk+\bfk')
2\sqrt{\Omega_{\alpha0}\Omega_{\beta0}}
\rho^{\alpha\kappa*}M_{\kappa\beta}(k)
\kma\eea
\bea\nonumber
&&<\tilde 0|a^\dagger_\alpha(\bfk)a_\beta(\bfk')|\tilde 0>=
\sqrt{\frac{\Omega_{\alpha0}\Omega_{\beta0}}
{\Omega_{\lambda0}\Omega_{\kappa0}}}\rho^{\alpha\kappa*} \rho^{\beta\lambda}<\tilde 0|a_\kappa(-\bfk)
a^\dagger_\lambda(-\bfk')|\tilde 0>
\\ \label{adaggeraexval}
&=&(2\pi)^3\delta^3(\bfk-\bfk')
2\sqrt{\Omega_{\alpha0}\Omega_{\beta0}}
\rho^{\alpha\kappa*}\rho^{\beta\lambda}M_{\kappa\lambda}(k)
\pkt\eea

In terms of the matrices $\rho$ and $M$ the fluctuation
integral, evaluated in the Bogoliubov-transformed vacuum takes the form
\bea\nonumber
&&\tilde \calf_{ij}(t)=
<\tilde 0|\psi_i(\bfx,t)\psi_j(\bfx,t)|\tilde 0>
\\\nonumber
&=&\int\frac{d^3 k}{(2\pi)^3}
\sum_{\alpha,\beta}
\frac{1}{2\sqrt{\Omega_{\alpha0}\Omega_{\beta0}}}
\left[f_i^\alpha(t)f_j^\beta(t)
\rho^{\beta\kappa}M_{\alpha\kappa}\right.
\\\nonumber
&&+f_i^\alpha(t)f_j^{\beta*}(t)
M_{\alpha\beta}
\\\nonumber
&&+f_i^{\alpha*}(t)f_j^{\beta}(t)
\rho^{\alpha\kappa*}\rho^{\beta\lambda}
M_{\kappa\lambda}
\\
&&\left.+ f_i^{\alpha*}(t)f_j^{\beta*}(t)
\rho^{\alpha\kappa*} 
M_{\kappa\beta} \right]
\pkt \eea
This is the basis for determining $\rho$, this is discussed 
in Sec. \ref{initialstate}.

As we have performed a canonical transformation it is to be expected
that $\tilde\calf_{ij}$ is real and symmetric in $i$ and $j$, as it
holds for $\calf_{ij}$. 
The sum of the  first and fourth terms in the bracket 
can be shown to be real and symmetric in
$i$ and $j$ using the relation \eqn{MrhorhoMT}, the symmetry of $\rho$ and
the Hermiticity of $M$.   
The sum of the second and third terms in the bracket can be 
rewritten, using Eq. \eqn{Meq} and the symmetry
in the summation over $\alpha$ and $\beta$, as
\bea\nonumber
&&M_{\alpha\beta}f^\alpha_if^{\beta*}_j
+\rho^{\alpha \lambda} M_{\kappa\lambda} 
\rho^{\beta\kappa*}f^{\beta*}_i f^\alpha_j
\\
&&=
\delta_{\alpha\beta}f^\alpha_if^{\beta*}_j+
\rho^{\alpha \lambda} M_{\kappa\lambda} 
\rho^{\beta\kappa*}f^\alpha_i f^{\beta*}_j+
\rho^{\alpha \lambda} M_{\kappa\lambda} 
\rho^{\beta\kappa*}f^{\beta* }_i f^\alpha_j 
\pkt\eea 
The first term on the right hand side  is the one that 
appears in the fluctuation integrals
$\calf_{ij}$. Its sum over $\alpha=\beta$ with 
prefactor $1/\Omega_{\alpha0}$ is real and
symmetric in $i$ and $j$, see
Eq. \eqn{ff}. The sum of the second and third terms on the right
hand side is
obviously symmetric in $i$ and $j$. It can be shown to be real
as well. 

The analogy of the various matrices we have defined here with the coefficients
obtained in the one-channel case of Ref. \cite{Baacke:1997zz} is given by 
$C \leftrightarrow \cosh \gamma$, 
$S\leftrightarrow \sinh \gamma \exp(i \delta)$,
$\rho \leftrightarrow\tanh \gamma \exp(i\delta)$,
 $M \leftrightarrow \cosh^2 \gamma$ , 
$M \rho \leftrightarrow \sinh 2\gamma \exp(i\delta)/2$,
and $2 M - I \leftrightarrow \cosh 2\gamma$.


\section{Canonical formalism at $t>0$}
\label{canform}
\setcounter{equation}{0} 
In Sec. \ref{model} we have defined the fluctuation integral
\bea \nonumber
\calf_{ij}(t)&=&< \psi_i(\bfx,t)\psi_j(\bfx,t) >
\\\nonumber
&=&\sum_\alpha
\int\frac{d^3k}{(2\pi)^3 2 \Omega_{\alpha0}(k)}
\left[f_i^\alpha(k,t)f_j^{\alpha*}(k,t)\right]
\pkt
\eea
The expression on the right hand side does not appear to be symmetric
in the indices $i$ and $j$, and does not appear to be real. On the
other hand the fields $ \psi_i(\bfx,t)$ and $\psi_j(\bfx,t)$
should commute with each other. Furthermore, the
commutator between  $ \psi_i(\bfx,t)$ and $\psi_j(\bfy,t)$ is given by the same
integral with the only modification that a factor $\exp[i\bfk(\bfy-\bfx)]$
appears in the integrand. As these fields commute as well for arbitrary
$\bfx$ and $\bfy$, the expression
\be \label{fluctintint}
\sum_\alpha
\frac{1}{ 2 \Omega_{\alpha0}(k)}
f_i^\alpha(k,t)f_j^{\alpha*}(k,t)
\ee
should be real. This is not quite obvious.

To begin with we consider the commutators in $\bfx$ space.
If we calculate the time derivative of the equal time commutator between
the fields we get
\be \label{ddtpsipsi}
\frac{d}{dt}[\psi_i(\bfx,t),\psi_j(\bfy,t)]=
[\psi_i(\bfx,t),\dot \psi_j(\bfy,t)]+
[\dot \psi_i(\bfx,t),\psi_j(\bfy,t)]
\kma\ee
and this is zero if the canonical commutation relations
\be\label{cancom}
[\psi_i(\bfx,t),\dot \psi_j(\bfy,t)]=i \delta_{ij}\delta(\bfx-\bfy)
\ee
hold at time $t$. If we require that this relation continues to hold 
we get the condition
\be\label{ddtpsidotpsi}
\frac{d}{dt}[\psi_i(\bfx,t),\dot \psi_j(\bfy,t)]= 
[\dot\psi_i(\bfx,t),\dot \psi_j(\bfy,t)]+
[\psi_i(\bfx,t),\ddot \psi_j(\bfy,t)]=0
\pkt\ee
The second term can be expressed, using the equation of motion
\be
\ddot \psi_j-\Delta \psi_j+\calm_{jk}\psi_k=0
\kma \ee
by the field commutators; the term vanishes if these commutators take
their canonical form at time $t$.
We have to require that the first term vanishes:
\be
[\dot\psi_i(\bfx,t),\dot \psi_j(\bfy,t)]=0
\ee
at time $t$. If this identity shall continue to hold we have
to make sure that
\be
\label{ddtdotpsidotpsi}
\frac{d}{dt}
[\dot\psi_i(\bfx,t),\dot \psi_j(\bfy,t)]=
[\dot\psi_i(\bfx,t),\ddot \psi_j(\bfy,t)]+
[\ddot\psi_i(\bfx,t),\dot \psi_j(\bfy,t)]=0
\pkt\ee
Using again the equation of motion and the symmetry of $\calm_{ij}(t)$
this can be verified, if the nontrivial commutations \eqn{cancom}
hold, whereupon the scheme closes.

This is of course the standard way for proving the time independence
of the canonical commutation relations. But this analysis in $\bfx$ space 
show us how to proceed in proving the identity
\be \label{ff}
\Im \sum_\alpha
\frac{1}{ \Omega_{\alpha0}(k)}
f_i^\alpha(k,t)f_j^{\alpha*}(k,t)=0
\ee 
that guarantees the reality and symmetry of the fluctuation
integrals. In order 
for the identity \eqn{ff} to hold at all times,
we have to require in addition that  the identities
\be\label{ffp}
\Im \sum_\alpha
\frac{1}{\Omega_{\alpha0}(k)}
f_i^\alpha(k,t)\dot f_j^{\alpha*}(k,t)=\delta_{ij}
\ee
and 
\be\label{fpfp}
\Im \sum_\alpha
\frac{1}{ \Omega_{\alpha0}(k)}
\dot f_i^\alpha(k,t)\dot f_j^{\alpha*}(k,t)=0
\ee
hold independent of time, {\em and that they hold at $t=0$}.
There is no direct evidence for any of these relations; we just can prove
that they continue to hold if they hold at one time. In the one-field 
case one just has to prove
that the canonical commutator is satisfied at all times and that
follows from the conservation of the Wronskian. Here the Wronskian of the
fluctuations is given, with our initial conditions, by
\be\label{wronskian}
W (f^\alpha,f^\beta)=\sum_i\left(f^\alpha_i\dot f^{\beta*}_i-
\dot f^\alpha_i f^{\beta*}_i\right)=2 i \delta_{\alpha\beta}
\Omega_{\alpha0}
\kma\ee
where the summation is with respect to the lower indices, while
in the commutators we need summations over the upper indices, weighted 
with $1/\Omega_{\alpha0}$.

The proof that the conditions Eqs. \eqn{ff}-\eqn{fpfp} hold independent
of time goes through in analogy to the proof in $\bfx$ space given
above, using this time the equations of motion \eqn{fdgl}.
The time derivative
(on both  sides) of Eq. \eqn{ff} holds, if the relation \eqn{ffp}
holds. The time derivative of Eq. \eqn{ffp} can be shown to
hold by using the equations of motion for the fluctuations, and assuming
that the relations \eqn{ff} and  \eqn{fpfp} hold. 
Using again the equations
of motion, the time derivative of the relation \eqn{fpfp} holds if Eq.
\eqn{ffp} holds.
 
We still have to consider the initial time $t=0$.
With the initial conditions Eqs. \eqn{finit} and \eqn{fpinit}
the relations \eqn{ff} and \eqn{fpfp} hold trivially as we have chosen
the $f_i^\alpha(0)=f_{i0}^\alpha$ to be the real eigenvectors 
of the mass matrix. We could still multiply the two eigenvectors with 
two different phase factors $\exp(i\delta_{\alpha})$ without spoiling these
conditions. Eq.\eqn{ffp} at $t=0$ reduces to
\be\label{ffp0}
\sum_\alpha
f_{i0}^\alpha f_{j0}^{\alpha}=\delta_{ij}
\kma\ee 
and this is the orthogonality relation dual to
\be
\sum_i
f_{i0}^\alpha f_{i0}^{\beta}=\delta_{\alpha\beta}
\pkt\ee
So at $t=0$ all three relations are satisfied, and then will
so for $t>0$.

Aside from their importance for the formalism developed here
the relations \eqn{ff}-\eqn{fpfp} represent useful checks
for  numerical simulations, along with the time independence
of the Wronskian;  we have verified this numerically.

\end{appendix}

\bibliography{bogopap}

\begin{thebibliography}{10}

\bibitem{Maslov:1997bf}
V.~P. Maslov and O.~Y. Shvedov,
\newblock Theor. Math. Phys. {\bf 114}, 184 (1998), [hep-th/9709151].

\bibitem{Baacke:1997zz}
J.~Baacke, K.~Heitmann and C.~Patzold,
\newblock Phys. Rev. {\bf D57}, 6398 (1998), [hep-th/9711144].

\bibitem{Anderson:2001th}
P.~R. Anderson, W.~Eaker, S.~Habib, C.~Molina-Paris and E.~Mottola,
\newblock Int. J. Theor. Phys. {\bf 40}, 2217 (2001).

\bibitem{Goldstein:2002fc}
K.~Goldstein and D.~A. Lowe,
\newblock Phys. Rev. {\bf D67}, 063502 (2003), [hep-th/0208167].

\bibitem{Danielsson:2002mb}
U.~H. Danielsson,
\newblock JHEP {\bf 12}, 025 (2002), [hep-th/0210058].

\bibitem{Burgess:2003hw}
C.~P. Burgess, J.~M. Cline, F.~Lemieux and R.~Holman,
\newblock astro-ph/0306236.

\bibitem{Schalm:2004qk}
K.~Schalm, G.~Shiu and J.~P. van~der Schaar,
\newblock JHEP {\bf 04}, 076 (2004), [hep-th/0401164].

\bibitem{Anderson:2005hi}
P.~R. Anderson, C.~Molina-Paris and E.~Mottola,
\newblock Phys. Rev. {\bf D72}, 043515 (2005), [hep-th/0504134].

\bibitem{Collins:2005nu}
H.~Collins and R.~Holman,
\newblock Phys. Rev. {\bf D71}, 085009 (2005), [hep-th/0501158].

\bibitem{Collins:2006bg}
H.~Collins and R.~Holman,
\newblock Phys. Rev. {\bf D74}, 045009 (2006), [hep-th/0605107].

\bibitem{Greene:2004np}
B.~R. Greene, K.~Schalm, G.~Shiu and J.~P. van~der Schaar,
\newblock JCAP {\bf 0502}, 001 (2005), [hep-th/0411217].

\bibitem{Borsanyi:2008ar}
S.~Borsanyi and U.~Reinosa,
\newblock 0809.0496.

\bibitem{Baacke:1997rs}
J.~Baacke, K.~Heitmann and C.~Patzold,
\newblock Phys. Rev. {\bf D56}, 6556 (1997), [hep-ph/9706274].

\bibitem{Baacke:1999nq}
J.~Baacke and C.~Patzold,
\newblock Phys. Rev. {\bf D62}, 084008 (2000), [hep-ph/9912505].

\bibitem{Stueckelberg:1951zz}
E.~C.~G. Stueckelberg,
\newblock Phys. Rev. {\bf 81}, 130 (1951).

\bibitem{Bogoliubov:1980}
N.~Bogoliubov and D.~Shirkov,
\newblock {\em Introduction to the Theory of Quantized Fields} (Wiley, New
  York, NY, 1980).

\bibitem{Symanzik:1981wd}
K.~Symanzik,
\newblock Nucl. Phys. {\bf B190}, 1 (1981).

\bibitem{Cooper:1987pt}
F.~Cooper and E.~Mottola,
\newblock Phys. Rev. {\bf D36}, 3114 (1987).

\bibitem{Baacke:1999ia}
J.~Baacke, D.~Boyanovsky and H.~J. de~Vega,
\newblock Phys. Rev. {\bf D63}, 045023 (2001), [hep-ph/9907337].

\bibitem{Baacke:1998di}
J.~Baacke, K.~Heitmann and C.~Patzold,
\newblock Phys. Rev. {\bf D58}, 125013 (1998), [hep-ph/9806205].

\bibitem{Baacke:1999gc}
J.~Baacke and C.~Patzold,
\newblock Phys. Rev. {\bf D61}, 024016 (1999), [hep-ph/9906417].

\bibitem{Baacke:2003bt}
J.~Baacke and A.~Heinen,
\newblock Phys. Rev. {\bf D69}, 083523 (2004), [hep-ph/0311282].

\bibitem{Baacke:1996se}
J.~Baacke, K.~Heitmann and C.~Patzold,
\newblock Phys. Rev. {\bf D55}, 2320 (1997), [hep-th/9608006].

\bibitem{Nilles:2001fg}
H.~P. Nilles, M.~Peloso and L.~Sorbo,
\newblock JHEP {\bf 04}, 004 (2001), [hep-th/0103202].

\bibitem{Linde:1993cn}
A.~D. Linde,
\newblock Phys. Rev. {\bf D49}, 748 (1994), [astro-ph/9307002].

\bibitem{Copeland:1994vg}
E.~J. Copeland, A.~R. Liddle, D.~H. Lyth, E.~D. Stewart and D.~Wands,
\newblock Phys. Rev. {\bf D49}, 6410 (1994), [astro-ph/9401011].

\bibitem{Garcia-Bellido:1997wm}
J.~Garcia-Bellido and A.~D. Linde,
\newblock Phys. Rev. {\bf D57}, 6075 (1998), [hep-ph/9711360].

\bibitem{Olive:2006uw}
K.~A. Olive and M.~Peloso,
\newblock Phys. Rev. {\bf D74}, 103514 (2006), [hep-ph/0608096].

\bibitem{Allahverdi:2006xh}
R.~Allahverdi and A.~Mazumdar,
\newblock JCAP {\bf 0708}, 023 (2007), [hep-ph/0608296].

\bibitem{Birrell:1982}
N.~Birrell and P.~Davies,
\newblock {\em Quantum fields in curved space} (Cambridge University Press,
  Cambridge, 1982).

\bibitem{Schwinger:1960qe}
J.~S. Schwinger,
\newblock J. Math. Phys. {\bf 2}, 407 (1961).

\bibitem{Keldysh:1964ud}
L.~V. Keldysh,
\newblock Zh. Eksp. Teor. Fiz. {\bf 47}, 1515 (1964),
\newblock [Sov. Phys. JETP {\bf 20}, 1018 (1965)].

\bibitem{Calzetta:1986ey}
E.~Calzetta and B.~L. Hu,
\newblock Phys. Rev. {\bf D35}, 495 (1987).

\bibitem{Jordan:1986ug}
R.~D. Jordan,
\newblock Phys. Rev. {\bf D33}, 444 (1986).

\bibitem{Ringwald:1987ui}
A.~Ringwald,
\newblock Ann. Phys. {\bf 177}, 129 (1987).

\bibitem{Boyanovsky:1992vi}
D.~Boyanovsky and H.~J. de~Vega,
\newblock Phys. Rev. {\bf D47}, 2343 (1993), [hep-th/9211044].

\bibitem{Polarski:1995jg}
D.~Polarski and A.~A. Starobinsky,
\newblock Class. Quant. Grav. {\bf 13}, 377 (1996), [gr-qc/9504030].

\bibitem{Khlebnikov:1996mc}
S.~Y. Khlebnikov and I.~I. Tkachev,
\newblock Phys. Rev. Lett. {\bf 77}, 219 (1996), [hep-ph/9603378].

\bibitem{Baacke:2009?}
J.~Baacke, L.~Covi and N.~Kevlishvili,
\newblock work in progress.

\bibitem{Bogoliubov:1983}
N.~Bogoliubov and D.~Shirkov,
\newblock {\em Quantum fields} (Benjamin/Cummings, Reading, MA, 1983).

\end{thebibliography}
\bibliographystyle{h-physrev4}

\end{document}